# Sub-ps thermionic electron injection effects on exciton-formation dynamics at a van der Waals semiconductor/metal interface

Kilian R. Keller[1,¶], Ricardo Rojas-Aedo[1,¶], Huiqin Zhang[2], Pirmin Schweizer[1], Jonas Allerbeck[1,3], Daniele Brida[1], Deep Jariwala[2], and Nicolò Maccaferri[1,4,*]

1. Department of Physics and Materials Science, University of Luxembourg, 162a avenue de la Faïencerie, L-1511 Luxembourg, Luxembourg
2. Department of Electrical and Systems Engineering, University of Pennsylvania, Philadelphia, PA, 19104, USA
3. Nanotech@surfaces Laboratory, EMPA, Ueberlandstrasse 129, 8600 Dübendorf, Switzerland
4. Department of Physics, Umeå University, Linnaeus väg 20, SE-90736 Umeå, Sweden

[¶]These authors contributed equally to this work

nicolo.maccaferri@umu.se

**Inorganic van der Waals bonded semiconductors like transition metal dichalcogenides are subject of intense research due to their electronic and optical properties which are promising for next-generation optoelectronic devices. In this context, understanding the ultrafast carrier dynamics, as well as charge and energy transfer at the interface between metals and semiconductors is crucial and yet quite unexplored. Here, we present an experimental study on how thermally induced ultrafast charge carrier injection affects the exciton formation dynamics in bulk WS2 by employing a pump-push-probe scheme, where a pump pulse induces thermionic injection of electrons from the gold substrate into the conduction band of the semiconductor, and another delayed push pulse excites direct transitions in the WS2. The transient response shows different dynamics on the sub-ps timescale by varying the delay between pump and push pulses or by changing the pump fluence, thus disclosing the important role of ultrafast hot electron injection on the exciton formation dynamics. Our findings might have potential impact on research fields that target the integration of ultrafast optics at the boundary of photonics and electronics, as well as in optically-driven CMOS and quantum technologies.**

Heterojunctions of metals and semiconducting transition metal dichalcogenides (TMDs) allow various possibilities for the manipulation and exploitation of light-matter interactions, such as the control of plasmonic excitations [1-4] and plasmon-induced charge injection [5-8], transistors [9], and photovoltaics [10]. Due to their layered structure, excited electrons and holes in TMDs exhibit enhanced Coulomb interactions in both monolayer and bulk (> 5 layers) forms, leading to room-temperature stable excitons, which dominate the optical and charge transport properties in these materials. Furthermore, TMDs form atomically clean and sharp interfaces with other materials [11], which makes them ideal candidates for optoelectronic applications where high-quality interfaces between metals and semiconductors are essential. Moreover, TMDs potentially offer a superior alternative to other semiconductors, as TMD/metal interfaces show weak Fermi-level pinning [12]. For these reasons, the exploitation of TMDs for optoelectronics is subject of current intense research [13] where different degrees of freedom, such as manipulation of the dielectric environment [14], and exciton-plasmon interaction [15], have been explored. As well, the ultrafast electronic dynamics of isolated 2D and bulk TMDs interfaced with insulating substrates have been the focus of recent studies [16-18].

Here, we show a new perspective to study the effect of charge injection on exciton dynamics in inorganic semiconductors in view of future applications which exploits the ultrafast (sub-ps) opto-electronic properties of TMDs. In more detail, we focus on how thermally induced ultrafast free charge carrier injection from a metal affects the exciton formation dynamics in a bulk TMD/metal heterojunction. For this purpose, we measure the ultrafast transient response of the heterojunction employing two different experimental schemes: standard pump-probe (PP) and a three-pulse pump-push-probe (PPP) configurations. The latter approach enables to disentangle the effect of hot-electrons injection from the metallic substrate, from the direct excitations in the semiconductor.

The TMD employed in our study is tungsten disulfide ($WS_2$), since it is a promising material for applications given its superior charge transport performance compared to other TMDs [19] and, most importantly, because it displays a single and very strong primary exciton feature which dominates the optical spectrum even in the bulk form and at room temperature. The A-exciton exhibits a binding energy of about 50 meV given an electronic band gap at the K-point of 2.1 eV [20,21] in bulk $WS_2$. We also chose a bulk sample of $WS_2$ instead of monolayer, due to higher absorption and lower contact resistance at the TMD/metal interface for charge injection [14]. After optical excitation, the ultrafast dynamics in inorganic semiconductors are dominated by carrier-carrier (c-c) scattering, that involves electron-electron and electron-hole scattering, leading to exciton formation, which typically happens on a timescale less than one picosecond [18]. In our study we observe these dynamics by exciting a free electron-hole plasma in the $WS_2$ with a narrowband laser pulse tuned at a wavelength of 515 nm (2.4 eV) in order to ensure that we do not directly excite the A-exciton. For the metal we employ gold, since it displays a large work function (WF) of approximately 5.1 eV thus leading to a lower Fermi level pinning effect and oxidation that otherwise would introduce additional resistance for injection [22]. The fact that gold has a WF that exceeds the electron affinity of $WS_2$ is of further importance because, in the reverse case, an accumulation layer for electrons would form at the $WS_2$/Au interface with a built-in field that impairs the injection of electrons into the semiconductor. In order to draw conclusions about the effect of the charge injection on the ultrafast electronic dynamics in the $WS_2$, we also measure as a reference a $WS_2$ sample deposited on a $SiO_2$ substrate.

The $WS_2$ sample is directly exfoliated on gold, leading to weak electronic coupling, which results in the formation of a Schottky junction [23] with distribution of metal electronic states and band bending in the $WS_2$ in proximity of the interface as sketched in Figure 1a (more information about sample preparation can be found in the Methods Section). In our case, an important parameter affecting the contact resistance is the so called Schottky barrier height (SBH), which is the potential barrier that the hot electrons have to overcome to be injected from the gold into the conduction band of the $WS_2$. For the $WS_2$/Au junction the SBH is approximately 1 eV [22]. Figure 1b depicts the conditions of the first experiment in which we sent a laser pulse centered at 515 nm (2.4 eV) to the sample that causes two main effects: (i) an increase of the electronic gas temperature in gold leads to thermionic injection of hot electrons into the conduction band of $WS_2$, and (ii) due to the photon energy of 2.4 eV this pulse causes a direct excitation of free electrons and holes in the $WS_2$.

Figure 1c shows the steady state spectra of $WS_2$/Au (green) and $WS_2$/$SiO_2$ (blue) in reflection and transmission, respectively. The dips at 618 nm (2.01 eV) for $WS_2$/Au and at 630 nm (1.97 eV) for $WS_2$/ $SiO_2$ correspond to the absorption of the A-exciton. The difference of the excitonic resonances can be attributed to a different screening from the gold at the $WS_2$/Au interface compared to the $WS_2$/$SiO_2$ sample. From the spectral position of the ethalon mode at 730 nm (1.70 eV) we can determine the thickness of the WS2 flake on gold [4], which is about 20 nm. The $WS_2$/$SiO_2$ sample has an approximate thickness of around 100 nm. The difference in thickness does not change the ultrafast electronic response in $WS_2$ [16], since both can be considered to be bulk. In the PP measurements, we focus on the neutral A-exciton absorption spectral range, which has an additional minor contribution from the negatively charged trion absorption at slightly lower energy with respect to the A-exciton. For this reason, we detect the probe signal by using a band-pass filter centered at 610 nm (2.03 eV) with a spectral width of 10 nm, depicted by the red bar in Figure 1c.

In the PP study on WS$_2$/Au and WS$_2$/SiO$_2$, we use an optical pulse centered at 515 nm (2.4 eV) with a pulse duration of about 150 fs and a fluence of 200 µJ/cm$^2$ to pump the system. As a probe pulse we use visible supercontinuum white light with a fluence of about 40 µJ/cm$^2$. We refer to this first PP study as our benchmark measurement throughout the manuscript. Figure 1d shows the transient response (ΔS/S) of WS$_2$/Au (green line), WS$_2$/SiO$_2$ (blue line) and bare gold (dashed orange line) as a function of PP delay $t_2$. The measurements show, that the presence of a Schottky interface leads to fundamentally different dynamics as can be seen as well by fitting the fast decay with a bi- and single exponential yielding $\tau_{WS_2/Au} = 211$ fs and $\tau_{WS_2/SiO_2} = 541$ fs for WS$_2$/Au and WS$_2$/SiO$_2$ respectively. A relevant observation here is that the measured dynamics of WS$_2$/SiO$_2$ agree with previous results on similar samples [17]. The difference between the curves cannot be explained just through the transient response of the gold since the signal from the bare gold alone is two orders of magnitudes lower. Therefore, it is likely that this change in dynamics is related to the free charges injected from the metal into the semiconductor, which via c-c scattering, screening and renormalization in combination with the screening effect from the semiconductor-metal interface, can modify both the SBH and the exciton binding energy. It is worth mentioning that the effect from the injection and the interface effects, i.e. screening of the WS$_2$ by the gold, are not strictly separable, as the injection changes the density of free electrons in the metal and that in turn modifies the interface itself. From Figure 1d, we see that there is a strong change of the dynamics for $t_2 < 1$ ps. As this temporal regime is dominated by effects like c-c scattering in the WS$_2$, we would expect a change of dynamics in this regime upon changes of the density of carriers either excited or injected. Therefore, we vary the fluence of the pump pulse and measure the transient absorption on the WS$_2$/Au. The normalized signals in Figure 1e show that the dynamics for $t_2 < 1$ ps are not changing upon variation of the pump fluence. For larger delays $t_2$ there is an offset for different fluences, which we attributed to heating of the system by the pump pulse to different equilibrium temperatures, that implies a different average number of phonons that modifies the electronic band-structure in WS$_2$ via electron-phonon (e-ph) scattering. To understand the results for $t_2 < 1$ ps, one first step is to remember the two main effects caused by the pump pulse. Changing the pump pulse fluence simultaneously modifies the density of injected carriers and the directly excited carriers in WS$_2$. In case of strong interaction between the injected carriers and the charge carriers in WS$_2$, it is reasonable to expect a significant change in the excited carrier dynamics when the ratio between the density of injected and excited carriers is varied. For this reason, it is necessary to control the injection independently from the excitation. More importantly, in order to understand how an ultrafast excitation in the WS$_2$ responds to an injection of carriers from the gold, the two processes have to be separated. This means that the thermionic injection from the gold and the carrier excitation in $WS_2$ should be generated by two different and independent laser pulses. For this reason, we performed a second study where we implemented a three-pulse PPP measurement scheme to detect the transient response of our heterojunction. Figure 2a depicts the outline of the PPP experiment. We refer to the first pulse arriving at the interface as "pump", using the fundamental wavelength of the laser amplifier at 1030 nm (1.2 eV), with an initial fluence of 1.7 mJ/cm$^2$.

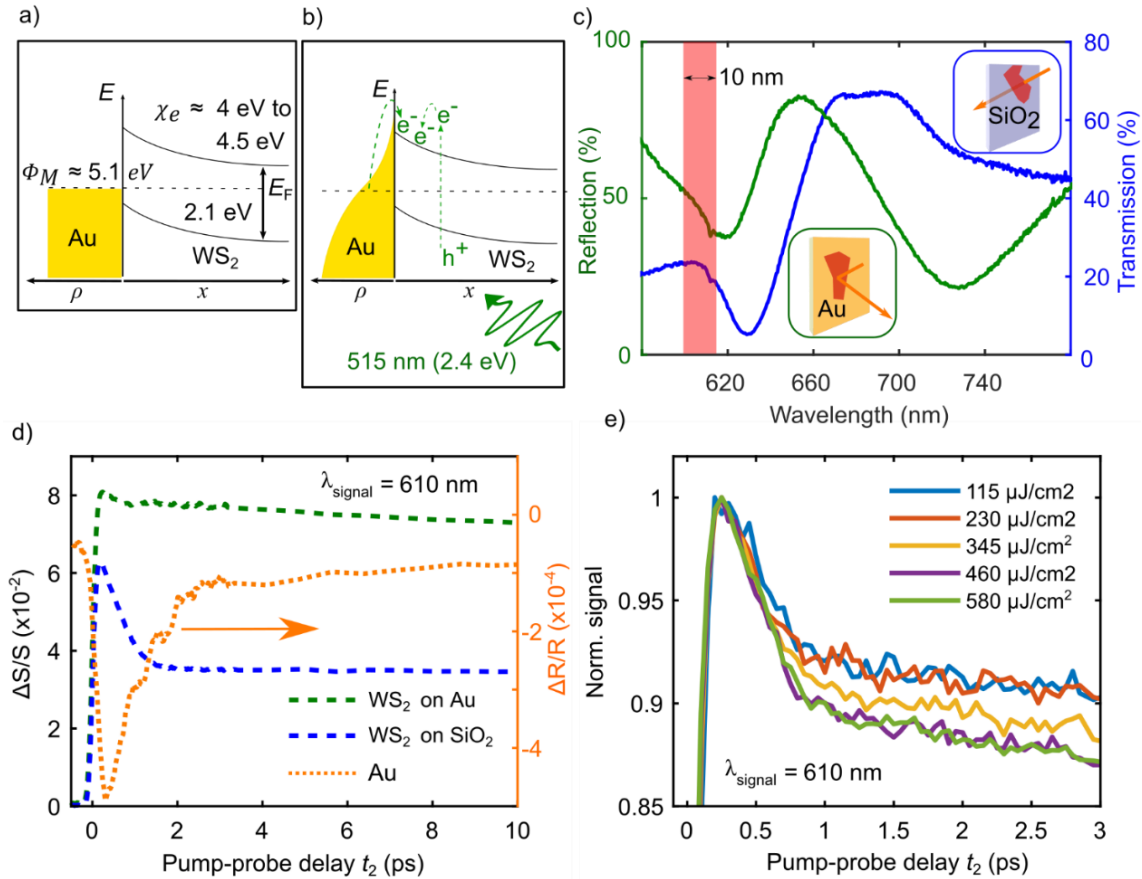

**Figure 1.** PP transient absorption measurement on $WS_2$/Au and $WS_2$/SiO$_2$ at $\lambda_{signal}$ = 610 nm (2.03 eV). a) Density of states ($\rho$) in gold and band alignment in $WS_2$ for $WS_2$/Au heterojunction with approximate values for gold work function $\Phi_M$, electron affinity $\chi_e$ of $WS_2$ and indication of Fermi energy level $E_F$. b) $\rho$ of gold and band alignment in case of illumination by laser pulse with indication of a direct excitation of free electrons (e$^-$) and holes (h$^+$) and thermionically injected electrons. c) Steady state spectra in reflection of $WS_2$/Au (green) and in transmission of $WS_2$/SiO$_2$ (blue). Red bar indicates spectral width of band pass centered at 610 nm (2.03 eV). d) PP measurements on $WS_2$/Au (green dashed line), $WS_2$/SiO$_2$ (blue dashed line) and bare gold substrate (orange dotted line). Pump at 515 nm (2.4 eV) with fluence of 200 μJ/cm$^2$ and visible white light probe with a fluence of 40 μJ/cm$^2$. ΔS/S represents either transient reflection ΔR/R for $WS_2$/Au or transient transmission ΔT/T for $WS_2$/SiO$_2$. d) Normalized ΔR/R PP measurement of $WS_2$/Au for different pump fluences.

The purpose of this first pulse is to increase the electronic temperature of gold and promote the thermionic injection of carriers into the WS$_2$. On the other hand, with a photon energy of 1.2 eV this pulse cannot directly excite carriers in the semiconductor. The subsequent pulse is called "push", and it is the previously used second harmonic at 515 nm (2.4 eV) with a fluence of 200 μJ/cm$^2$. Thus, it has sufficient photon energy to excite an electron-hole plasma in the WS$_2$. The system is then probed in the same way as in the PP experiment. The introduction of an additional pump pulse with photon energy below the electronic bandgap of the WS$_2$ is the essential part of our work, allowing us to largely separate the hot-electron injection from the gold from the carrier excitation in the WS$_2$. Furthermore, PPP enables us to change the ratio between injected and excited charges by changing the fluence of the respective pulses. It is important to note that changes of this ratio imply that we can explore a different environment for c-c scattering in WS$_2$ which would affect the dynamics of processes occurring on the timescale < 1 ps, e.g. exciton formation. Compared to the more conventional PP scheme, the PPP configuration enables to study carrier dynamics in the WS$_2$ system in contact with "hot" gold or, in other words, with a hot-electron reservoir. Figure 2b summarizes the PPP measurement scheme, in which the pump pulse arrives at a fixed delay $t_1$ before the modulated push pulse, which is followed by the probe pulse with variable delay $t_2$. Heating of the gold due to the push pulse can be neglected since the fluence of the pump pulse is about an order of magnitude higher. In Figure 2c we plot the result of a PPP measurement on WS$_2$/Au (red line) and on the bare gold substrate (orange dotted line), for the case $t_1 = 0$ ps (red line) when pump and push pulse arrive at the same time, and the benchmark measurement (green dashed line). By comparing the PPP on WS$_2$/Au with the benchmark, it is evident that by adding the pump pulse in the former the dynamics for $t_2 < 1$ ps are qualitatively different, as the time constant associated with the fast decay component seems to become shorter in the PPP configuration. The fact that c-c scattering is the dominant effect for short delays $t_2$ implies that the pump in PPP introduces a different environment for scattering in the WS$_2$ by altering the injected to excited carrier ratio. As in the case of the PP study, we can see from the PPP measurements on the bare gold that the contribution from the metal alone is two orders of magnitude weaker, and thus cannot explain this difference in dynamics.

To better understand the change of dynamics caused by the hot-electron injection due to the pump pulse, we performed PPP measurements for different pump-push delays ($t_1$) on the WS$_2$/Au sample, and also compared the results with those observed in the case of the WS$_2$/SiO$_2$ reference sample. Figure 3a shows the PPP measurement at 610 nm (2.03 eV) on the WS$_2$/Au sample for the cases when the pump arrives 0 ps (red line), 0.1 ps (blue line) and 0.2 ps (brown line) before the push pulse.

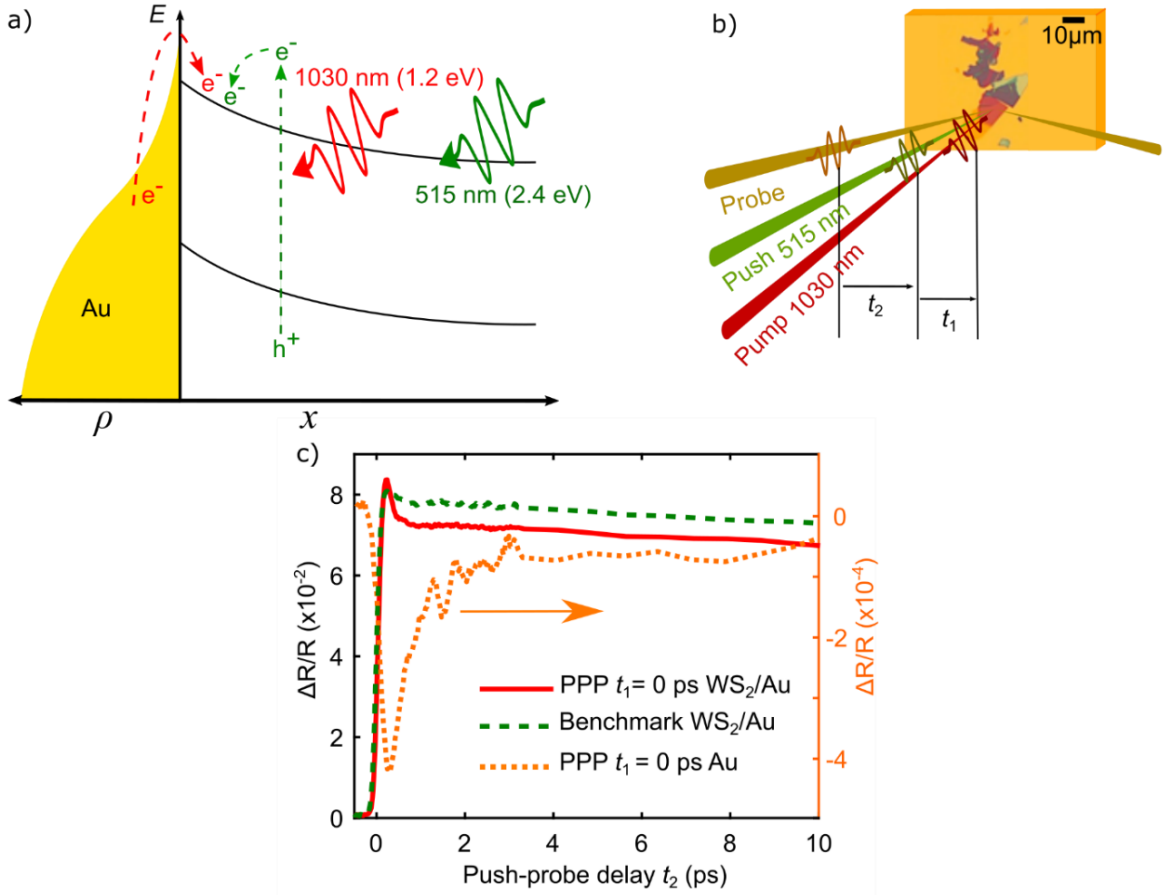

**Figure 2.** PPP experiment on $WS_2/Au$ at $\lambda_{signal} = 610$ nm (2.03 eV). a) Density of states ($\rho$) in gold and band alignment in $WS_2$ for $WS_2/Au$ heterojunction under illumination by pump pulse (red) followed by a modulated push (green) and probe pulse (brown). The pump induced thermionic injection of electrons ($e^-$) from the gold and the direct excitation of free electrons ($e^-$) and holes ($h^+$) in $WS_2$ is indicated by arrows. b) PPP configuration and microscopy image of $WS_2$ flakes on gold with indication of fixed pump–push delay $t_1$, scanned push-probe delay $t_2$ and fluences. c) PPP measurement with pump-push delay $t_1 = 0$ ps on $WS_2/Au$ (red line) and bare gold (orange dotted line) and reference PP measurement on $WS_2/Au$ (green dashed line).

The PPP curves show qualitatively strong variation in dynamics for delays $t_2 < 0.5$ ps (grey dotted line). For longer delays $t_2$ the curves follow the same dynamics with a constant offset with respect to the benchmark (green dashed line) of about $\Delta R/R \approx 0.5 \cdot 10^{-2}$. As mentioned in the discussion of Figures 1d and 2c, two temporal regimes must be distinguished: (i) the one dominated by c-c interaction (regime I) and (ii) the one dominated by thermal effects (regime II), that can be identified with respect to $t_2$. In the former, for $t_2 < 0.5$ ps the PPP for $t_1 = 0$ ps shows significantly different dynamics with respect to the benchmark, featuring a faster decay after the maximum of the $\Delta R/R$ signal. With increasing pump-push delay $t_1$, the dynamics approach the benchmark case recovering the same fast decay value for $t_1 = 0.2$ ps.

In Figure 3b we show the results of the PPP measurement at 610 nm (2.03 eV) on the $WS_2/SiO_2$ reference sample for different $t_1$. Similar to the $WS_2/Au$ sample, the experimental curves display different dynamics in regime I ($t_2 < 0.5$ ps) (grey dotted line), which for long $t_2$ delay converge and exhibit a comparable offset with respect to the benchmark of $\Delta T/T \approx 0.4 \cdot 10^{-2}$, like in the $WS_2/Au$ case. This similarity in regime II ($t_2 > 0.5$ ps) is reasonable, as the charge injection from the gold substrate is mostly affecting the short $t_2$ delays, and the effect of the pump in PPP for longer $t_2$ delays, i.e. heating of the system to different equilibrium temperatures, is similar for $WS_2/Au$ and $WS_2/SiO_2$. As for the first regime, the variation of the dynamics for $WS_2/SiO_2$ seems qualitatively smaller upon changing the pump-push delay. Furthermore, the dynamics of the benchmark are retrieved going to longer pump-push delays i.e around $t_1 = 5.1$ ps (pink curve) in case of the sample on glass. In order to better compare the evolution of the fast decay dynamics in this first regime ($t_2 < 0.5$ ps) between $WS_2/Au$ and $WS_2/SiO_2$, the experimental transient curves in Figure 3a and 3b have been fitted using a bi- and single exponential, respectively, to extract a time constant for this fast decay. Figure 3c shows the extracted fast decay times for the $WS_2$ on gold (green) and on glass (blue) for the different $t_1$. Values for the fast decay in $WS_2/Au$ are between three to four times shorter than for the sample on glass. The retrieval of the benchmark dynamics for increasing pump-push delays, which has a time constant of $\tau_{WS_2/Au} = 211$ fs and $\tau_{WS_2/SiO_2} = 541$ fs, happens for $t_1 < 1$ ps in case of $WS_2/Au$ and after several picoseconds pump-push delay in case of $WS_2/SiO_2$. Additionally, as shown in Fig. 3c, the variation of the fast decay time $\Delta$ amounts to $\Delta_{Au} \approx 104$ fs in $WS_2/Au$ compared to $\Delta_{SiO_2} \approx 46$ fs in $WS_2/SiO_2$ by changing $t_1$ in the sub-ps range. Thus, the observed differences in dynamics at short time delay $t_2$ can only be attributed to the effect of charge injection at the Schottky interface. The strong variation $\Delta_{Au}$ which results from changing the pump-push delay in steps of hundreds of femtoseconds corroborates the injection as responsible mechanism. Due to the thermal relaxation of the electrons in the metal and the ones injected in the semiconductor, the electrons undergo a temporal evolution through different thermal distributions. By changing the delay $t_1$, the initial thermal distribution of electrons is different at the moment of excitation in $WS_2$ induced by the push pulse.

To further investigate how the hot electrons affect the dynamics which take place within the time scale of regime I ($t_1 < 0.5$ ps), in the subsequent PPP measurements, we vary the push and pump fluences. We focused on the fixed pump-push delay of $t_1 = 0.1$ ps to avoid undesired effects at $t_1 = 0$ ps due to the temporal overlap of pump and push pulse, i.e excited-state absorption or nonlinear frequency mixing. Figure 4a shows the normalized $\Delta R/R$ signal at 610 nm (2.03 eV) obtained on $WS_2/Au$ for push fluences ranging from 200 μJ/cm^2 (blue line) to 30 μJ/cm^2 (yellow line) and the benchmark experiment (green dashed line). The curves show qualitatively no significant variation of the fast decay for short delays $t_2$, however for longer delays the PPP feature an offset with respect to the benchmark with a slight variation for the different push fluences. In this temporal regime, we attribute the offset and its variation to different lattice temperatures and e-ph scattering induced by the push pulse. Figure 4b shows the zoom on the raising part of the measurements shown in Figure 4a.

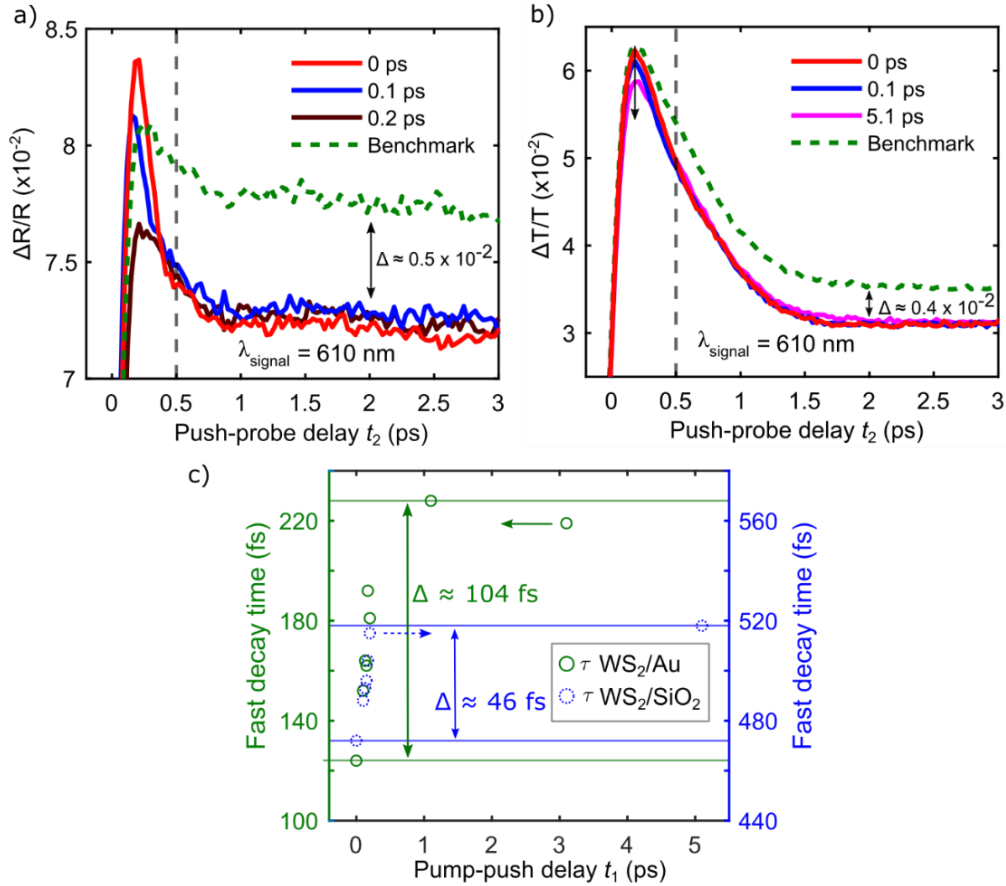

**Figure 3.** PPP on a) $WS_2/Au$ and b) $WS_2/SiO_2$ at $\lambda_{signal}$ = 610 nm (2.03 eV) for different pump-push delays $t_1$ and PP reference (green dashed line). The grey dotted line indicates the delay $t_2$ from which on different $t_1$ curves follow the same dynamics. c) Extracted time constants for fast decay of PPP measurements in a) and b) from bi- and single-exponential fits for $WS_2/Au$ (green circle) and $WS_2/SiO_2$ (blue dotted circle) respectively.

The qualitative independence from the push fluence observed also in the buildup dynamics of the ΔR/R signal strengthens our hypothesis that the dynamics for short delays $t_2$ are not affected by this parameter. If instead the ratio between injected and excited carriers is changed, we expect a change of dynamics in the first c-c interaction dominated temporal regime. To verify this dependence we also varied the pump pulse fluence in the PPP experiments, shown in Figure 4c. The highest fluence of 1.47 mJ/cm$^2$ exhibits a significantly faster decay for short delays $t_2$ compared to the benchmark, where the dynamics of the latter are recovered for decreasing pump fluences.

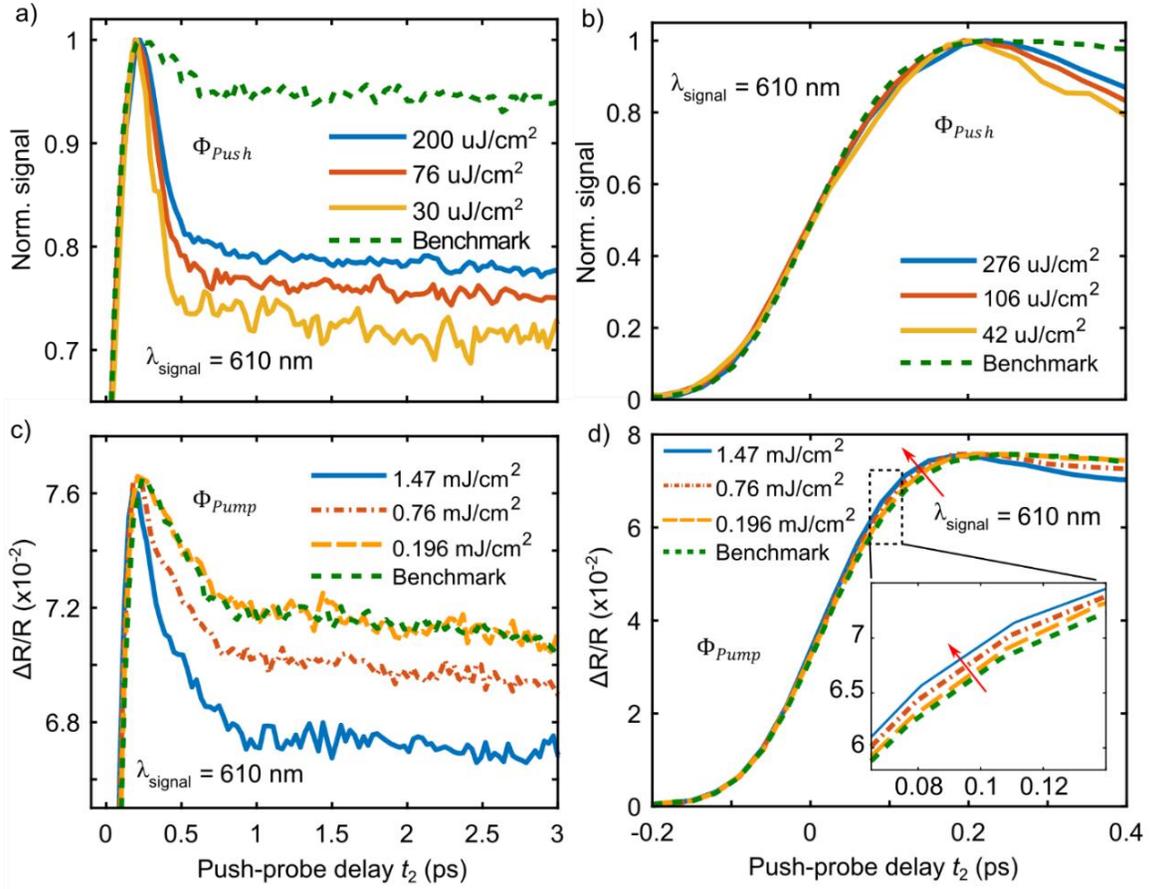

**Figure 4.** PPP curves at fixed pump-push delay $t_1$ = 0.1 ps at $\lambda_{signal}$ = 610 nm (2.03 eV) on WS$_2$/Au for different push ($\Phi_{Push}$) (a,b) and pump ($\Phi_{Pump}$) (c,d) fluences. Right panels show the build up dynamics of the measurements in a) and c). The inset in d) highlights the dynamics with red arrow indicating steepening of rise dynamics with increasing pump fluence.

Figure 4d shows the zoom on the dynamics of the raising part of the measurements in Figure 4c. We observe that the build-up dynamics display a systematic reduction in the rise time upon increase of the pump fluence. The variation of the pump fluence establishes different electronic temperatures in gold and consequently modifies the injected electron density. Therefore, the dependence of the fast decay and rise dynamics on the pump fluence, gives strong evidence for a change of the c-c scattering rate in the WS$_2$ and its effect on the observed dynamics.

In conclusion, we explored via PPP how thermionic hot electron injection at a WS$_2$/Au interface affects the reflection of a transient signal associated with the A-exciton dynamics in the semiconductor. By changing the pump-push delay or the fluence of the pump pulse, we observe different dynamics in the WS$_2$. Thus, we conclude that the charge injection from the gold to the WS$_2$, and consequently the ratio between injected carriers from the gold and the excited carriers in the WS$_2$, affects the ultrafast dynamics in WS$_2$. Our results show that the injected charged plasma only impacts the dynamics on the sub-ps timescale after excitation of the WS$_2$. In this temporal regime, the effect of injected carriers promotes a change in the rate of c-c scattering in WS$_2$ and consequently modifies the screening of the dielectric environment and the probability to form charged excitons i.e. trions, affecting intrinsically the exciton formation dynamics. So far transient absorption PP studies on isolated TMDs have shown a dependence of the ultrafast dynamical parameters such as the rise time constant in a transient absorption curve on the pump photon energy and an independence from the pump intensity [18]. The PPP scheme in this work demonstrates how these dynamics associated with the A-exciton of WS$_2$ that forms an interface with gold is affected by changing the fluence of a laser pulse instead of the photon energy. Our findings introduce an alternative approach to couple optoelectronic properties of a TMD/metal interface and exciton dynamics through electron injection across the Schottky barrier induced by an optical pulse. We foresee potential impact on research fields that target the integration of ultrafast optics at the boundary of photonics and electronics, such as in CMOS and quantum technology.

## METHODS

**Experiments.** We used the second harmonic of a Yb:KGW amplified laser modulated at 50 kHz and centered at 515 nm (2.4 eV), with a pulse duration of 150 fs as a pump(push) pulse in the PP(PPP) measurements. The pulse duration of the fundamental (pump pulse in the PPP measurements) at 1030 nm (1.2 eV) was 220 fs. As a probe pulse we used visible white light generated by the fundamental laser pulses inside a YAG crystal. Due to the narrowband detection, temporal compression of the probe pulse is not necessary. The temporal overlap between pump and probe $t_2 = 0$ is defined as the time when the normalized signal is equal to 0.5. The temporal overlap between pump and push $t_1 = 0$ was determined by generating a nonlinear optical signal between the two pulses. Modulation of the exciting pulse (515 nm) was achieved with a pockel cell. The sensitivity of our pump-probe setup allows to detect a variation of the transient signal given a root-mean-square of the noise floor on the order $10^{-4}$ to $10^{-5}$.

**Sample fabrication.** To ensure a flat surface of the gold back reflector rather than the rough surface of the evaporated Au, an epoxy-based peeling procedure was applied to the 100 nm-thick Au film evaporated on a clean polished Si wafer (parent wafer) using an e-beam evaporator (Kurt J. Lesker PVD 75). A piece of silicon wafer (transfer wafer) was glued to the Au film using a thin layer of thermal epoxy (Epo-Tek 375, Epoxy Technology) and then peeled upwards after the epoxy layer achieves its final hardness (curing), resulting in stripping of the Au film from the parent wafer. $WS_2$ was mechanically exfoliated from bulk crystal (HQ-graphene) using Scotch Tape and transferred onto the Au substrate.

## ACKNOWLEDGEMENTS

N.M. and D.B. acknowledge support from the Luxembourg National Research Fund (Grant No. C19/MS/13624497 'ULTRON') and the European Union under the FETOPEN-01-2018-2019-2020 call (Grant No. 964363 'ProID'). D.B. and K.R.K. acknowledges support from the European Research Council through grant no. 819871 (UpTEMPO). D.B. and R.R.-A. acknowledge support from ERDF Program (Grant No. 2017-03-022-19 'Lux-Ultra-Fast'). N.M. acknowledges support from the Swedish Research Council (Grant No. 2021-05784). D.J. and H.Z. acknowledge primary support from the U.S. Army Research Office under contract number W911NF-19-1-0109 as well as partial support from National Science Foundation supported University of Pennsylvania Materials Research Science and Engineering Center (MRSEC)(DMR-1720530). H.Z. also acknowledges support from Vagelos Institute of Energy Science and Technology graduate fellowship at the University of Pennsylvania. The Authors acknowledge Prof. Stefano Corni for fruitful discussions.

## REFERENCES

[1] J. Vogelsang, L. Wittenbecher, D. Pan, J. Sun, S. Mikaelsson, C. L. Arnold, A. L'Huillier, H. Xu, and A. Mikkelsen, "Coherent Excitation and Control of Plasmons on Gold Using Two-Dimensional Transition Metal Dichalcogenides," ACS Photonics **8**(6), 1607–1615 (2021).

[2] F. Davoodi and N. Talebi, "Plasmon-exciton Interactions in Gold-WSe2 Multilayer structures," ACS Appl. Nano Mater. **4**(6), 6067–6074 (2021).

[3] U. Celano and N. Maccaferri, "Chasing Plasmons in Flatland," Nano Lett. **19**(11), 7549–7552 (2019).

[4] H. Zhang, B. Abhiraman, Q. Zhang, J. Miao, K. Jo, S. Roccasecca, M. W. Knight, A. R. Davoyan, and D. Jariwala, "Hybrid exciton-plasmon-polaritons in van der Waals semiconductor gratings," Nat. Commun. **11**(1), 3552 (2020).

[5] Y. Kang, S. Najmaei, Z. Liu, Y. Bao, Y. Wang, X. Zhu, N. J. Halas, P. Nordlander, P. M. Ajayan, J. Lou, and Z. Fang, "Plasmonic Hot Electron Induced Structural Phase Transition in a MoS2 Monolayer," Adv. Mater. **26**(37), 6467–6471 (2014).

[6] L. Wang, Z. Wang, H. Y. Wang, G. Grinblat, Y. L. Huang, D. Wang, X. H. Ye, X. Bin Li, Q. Bao, A. S. Wee, S. A. Maier, Q. D. Chen, M. L. Zhong, C. W. Qiu, and H. B. Sun, "Slow cooling and efficient extraction of C-exciton hot carriers in MoS2 monolayer," Nat. Commun. **8**, 13906 (2017).

[7] A. O. Govorov, G. W. Bryant, W. Zhang, T. Skeini, J. Lee, N. A. Kotov, J. M. Slocik, and R. R. Naik, "Exciton-plasmon interaction and hybrid excitons in semiconductor-metal nanoparticle assemblies," Nano Lett. **6**(5), 984–994 (2006).

[8] H. S. Lee, D. H. Luong, M. S. Kim, Y. Jin, H. Kim, S. Yun, and Y. H. Lee, "Reconfigurable exciton-plasmon interconversion for nanophotonic circuits," Nat. Commun. **7**, 13663 (2016).

[9] B. Liu, Y. Ma, A. Zhang, L. Chen, A. Abbas, Y. Liu, C. Shen, H. Wan, and C. Zhou, "High-Performance WSe2 Field-Effect Transistors via Controlled Formation of In-Plane Heterojunctions," ACS Nano **10**(5), 5153–5160 (2016).

[10] J. Wong, D. Jariwala, G. Tagliabue, K. Tat, A. R. Davoyan, M. C. Sherrott, and H. A. Atwater, "High Photovoltaic Quantum Efficiency in Ultrathin van der Waals Heterostructures," ACS Nano **11**(7), 7230–7240 (2017).

[11] A. K. Geim and I. V. Grigorieva, "Van der Waals heterostructures," Nature **499**, 419–425 (2013).

[12] J. R. Lince, D. J. Carré, and P. D. Fleischauer, "Schottky-barrier formation on a covalent semiconductor without Fermi-level pinning: The metal-MoS2(0001) interface," Phys. Rev. B **36**(3), 1647–1656 (1987).

[13] T. LaMountain, E. J. Lenferink, Y. J. Chen, T. K. Stanev, and N. P. Stern, "Environmental engineering of transition metal dichalcogenide optoelectronics," Front. Phys. **13**(4), (2018).

[14] Z. Li, G. Ezhilarasu, I. Chatzakis, R. Dhall, C. C. Chen, and S. B. Cronin, "Indirect Band Gap Emission by Hot Electron Injection in Metal/MoS2 and Metal/WSe2 Heterojunctions," Nano Lett. **15**(6), 3977–3982 (2015).

[15] Y. H. Chen, R. R. Tamming, K. Chen, Z. Zhang, F. Liu, Y. Zhang, J. M. Hodgkiss, R. J. Blaikie, B. Ding, and M. Qiu, "Bandgap control in two-dimensional semiconductors via coherent doping of plasmonic hot electrons," Nat. Commun. **12**, 4332 (2021).

[16] J. He, D. He, Y. Wang, Q. Cui, F. Ceballos, and H. Zhao, "Spatiotemporal dynamics of excitons in monolayer and bulk WS2," Nanoscale **7**(21), 9526–9531 (2015).

[17] F. Ceballos, Q. Cui, M. Z. Bellus, and H. Zhao, "Exciton Formation in Monolayer Transition Metal Dichalcogenides," Nanoscale **8**(22), 11681–11688 (2016).

[18] C. Trovatello, F. Katsch, N. J. Borys, M. Selig, K. Yao, R. Borrego-varillas, F. Scotognella, I. Kriegel, A. Yan, A. Zettl, P. J. Schuck, A. Knorr, G. Cerullo, and S. D. Conte, "The ultrafast onset of exciton formation in 2D semiconductors," Nat. Commun. **11**, 5277 (2020).

[19] Z. Jin, X. Li, J. T. Mullen, and K. W. Kim, "Intrinsic transport properties of electrons and holes in monolayer transition-metal dichalcogenides," Phys. Rev. B **90**, 045422 (2014).

[20] A. Chernikov, T. C. Berkelbach, H. M. Hill, A. Rigosi, Y. Li, O. B. Aslan, D. R. Reichman, M. S. Hybertsen, and T. F. Heinz, "Exciton binding energy and nonhydrogenic Rydberg series in monolayer WS2," Phys. Rev. Lett. **113**, 076802 (2014).

[21] B. Zhu, X. Chen, and X. Cui, "Exciton binding energy of monolayer WS2," Sci. Rep. **5**, 9218 (2015).

[22] J. R. Durán Retamal, D. Periyanagounder, J. J. Ke, M. L. Tsai, and J. H. He, "Charge carrier injection and transport engineering in two-dimensional transition metal dichalcogenides," Chem. Sci. **9**(40), 7727–7745 (2018).

[23] K. Jo, P. Kumar, J. Orr, S. B. Anantharaman, J. Miao, M. J. Motala, A. Bandyopadhyay, K. Kisslinger, C. Muratore, V. B. Shenoy, E. A. Stach, N. R. Glavin, and D. Jariwala, "Direct Optoelectronic Imaging of 2D Semiconductor-3D Metal Buried Interfaces," ACS Nano **15**(3), 5618–5630 (2021).